# Functionalities of $Au_2O$, $Au_2O_3$, $Au_2O_{3-x}$, and nanosheets, including spontaneous polarization, using DFT and hybrid functional

Yukio Watanabe*

Kyushu University, Shiobara, Minami-ku, Fukuoka, Japan 815-8540
University of Hyogo, Shosya, Himeji, Hyogo, Japan 671-2201

We used density functional theories (DFT) to investigate the properties of $Au_2O$ and $Au_2O_{3-x}$ ($x$ = 0–0.08) to reveal their remarkable functionalities. Hybrid functional theories accurately estimate the band gap ($E_g$) of oxides, and the present hybrid functional calculations determined $E_g$ values of 0.96 eV for $Au_2O$ and 2.86 eV for $Au_2O_3$, which is >300% of the commonly accepted $E_g$ of $Au_2O_3$ (0.85 eV). Moreover, we discovered spontaneous polarization ($P_S$) in $Au_2O_3$, which is unusually large and advantageous for catalysis. The $P_S$ was retained even in 2-nm-thick $Au_2O_3$ nanosheets, similar to hyperferroelectric, generating a potential of 0.6 eV despite screening caused by surface reconstruction, which is a novel screening mechanism. Below a thickness of 0.8 nm, the $P_S$ vanished, and inversion symmetry emerged at 0.4 nm, suggesting a new approach to finding a paraelectric phase. $Au_2O$ was supersoft under shear distortion.

# I. Introduction

Among noble metals, Au is widely recognized for its chemical inertness and exceptional resistance to oxidation. This property makes Au suitable for diverse applications, such as in semiconductors [1], biomedical devices [2], and electrochemical systems [3]. However, in certain electrochemical reactions, the surfaces of Au anodes undergo oxidation [4]. Additionally, during manufacturing processes, Au-deposited materials are often annealed in oxidizing environments. Similar oxidation of Au surfaces has also been observed in plasma [5], ozone [6], and high-temperature $O_2$ atmospheres [7]. Upon oxidation, Au surfaces exhibit notable catalytic activity, and the essential role of oxygen in this behavior has been well documented [8].

To elucidate the properties of oxidized Au surfaces, oxygen-chemisorbed Au surfaces [6–9] and Au oxides have been extensively investigated. However, only limited studies have focused on bulk Au oxides [6,10–20]. In particular, these studies have examined the $Au^{1+}$, $Au^{2+}$, $Au^{3+}$, and $Au^{4+}$ oxidation states of bulk Au–O compounds. Of these, only $Au_2O_3$ and $Au_2O$ have been synthesized experimentally in bulk form [10–13]. Specifically, $Au_2O_3$ was synthesized under hydrothermal conditions by Schwarzmann et al. [10] and Weiher et al. [13]. Schwarzmann et al. [11] also synthesized single crystals of $Au_2O_3$, while Jones et al. [12] subsequently determined their crystal structures. Higo et al. [14] showed that gold surfaces became $Au_2O_3$ after the exposure to an oxygen glow discharge. Irissou et al. [15] reported that their vacuum-deposited Au oxide films consisted of $Au_2O_3$. Further, Pireaux et al. [6] observed that heating $Au_2O_3$ to approximately 450 K initiated decomposition, yielding $Au_2O$ as a breakdown product.

Beyond experimental studies, theoretical investigations have elucidated the structural and electronic properties of $Au_2O_3$ and $Au_2O$ and predicted those of their pure stoichiometric phases. For instance, Joshi et al. [16] and Tang et al. [17] used density functional theory (DFT) and the force field method to demonstrate that $Au_2O_3$ is stable under the pressure of 0–50 GPa, whereas $Au_2O$ can metastably exist under the pressure of 25–50 GPa. Filippetti and Fiorentini [18] observed that, according to DFT calculations, the cubic phase of $Au_2O$ was more stable than its face-centered cubic counterpart. Zhang et al. [19] and Tang et al. [17] predicted the existence and structure of hypothetical $AuO_4$ using DFT, while Hermann et al. [20] investigated the structural, electronic, and dynamic properties of hypothetical AuO using hybrid functional theory [21, 22]. Further, a DFT study conducted by Ren et al. [23] clarified the mechanism underlying the selective oxidation of vinyl chloride on $Au_2O$ (100) surfaces.

Additionally, Pei et al. [24] reported the optical properties, including the dielectric functions, of $Cu_2O$, $Ag_2O$, and $Au_2O$ using DFT. Using the generalized gradient approximation (GGA), Pei et al. [24] and Shi et al. [25] estimated the band gaps of $Au_2O$ and $Au_2O_3$, which were small and have been used as reference values. However, DFT calculations, including those performed using GGA, are known to inadequately account for electron correlation effects and fail to eliminate electron self-interactions, leading to inaccurate electronic and magnetic properties and erroneous band gap predictions. These deficiencies are effectively resolved by hybrid functionals, as demonstrated by numerous studies [21, 22, 26–30].

Meanwhile, spontaneous polarization ($P_S$) has been shown to facilitate efficient photocatalytic water splitting by spatially separating photogenerated electrons ($e^-$) and holes ($h^+$) [31, 32], while magnetism and spin ordering are thought to enhance catalytic activity [33].

Although bulk forms of Au oxides have been experimentally observed only in $Au_2O_3$ and $Au_2O$ [10–13], with $Au_2O_3$ considered the most important, the theoretical investigations of their properties have been limited to a few. Hence, we investigate their structural, dielectric, mechanical, electronic, and magnetic properties using DFT and hybrid functionals. These calculations reveal a high degree of functionality in these oxides, including pronounced $P_S$ and hyperferroelectric-like behavior in $Au_2O_3$. Furthermore, the hybrid functional calculations demonstrate that $Au_2O_3$ is an insulator with a band gap ($E_g$) sufficiently large to support dielectric and optical applications.

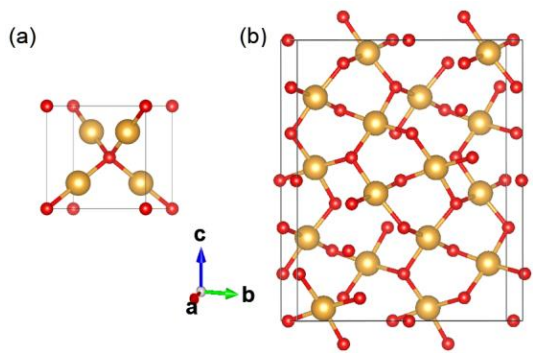

**Fig. 1.** Atomic positions in (a) $Au_2O$ and (b) $Au_2O_3$ unit cells. Au and oxygen atoms are represented by large yellow and small red circles, respectively. This color scheme is maintained in all subsequent figures.

**II. Methods**

Au$_2$O has a cuprite structure (cubic, Pn-3m symmetry), and its unit cell contains four Au and two O atoms [18, 23, 24]. Au$_2$O$_3$ is orthorhombic (Fdd2 symmetry), and its unit cell consists of 16 Au and 24 O atoms [12] (Fig. 1). DFT(GGA) and hybrid functional calculations used PAW potentials [34] implemented in VASP [35–37]. The plane wave energy cutoff was 650 eV, and a Gaussian function with a temperature broadening ($\sigma$) of 0.05 eV. We used Perdew-Burke-Ernzerhof functional optimized for solids (PBEsol) [38] as GGA and the Heyd‐Scuseria‐Ernzerhof functional for solids (HSEsol) [21] as the hybrid functional. The screening parameter ($\mu$) in HSEsol is the default value (0.2 Å$^{-1}$), and the *k*-grid reduction option of the Hartree-Fock kernel (Nkred) for reducing computational cost was not used in calculations of bulk Au$_2$O$_3$ and Au$_2$O. The PAW potentials were those recommended by the VASP manual for solid metal oxides: Au PAW potential for Au and O_s PAW potential for O in oxides with a cutoff energy of 282.85 eV.

The Monkhorst-Pack *k*-mesh for Brillouin-zone integration [39] in the geometry relaxation was 8×8×8 for Au$_2$O and 7×3×3 for Au$_2$O$_3$. All the ion positions were relaxed without restriction, and all the calculated forces were lower than 3 meV/Å after relaxation. $P_S$ was calculated using a Berry phase method [40] as implemented in VASP. The equilibrium lattice parameters calculated using PBEsol are often different from those using HSEsol [27, 29, 41]. However, in the case of Au oxides, such a difference was negligible; the equilibrium lattice constants of Au$_2$O calculated using PBEsol and HSEsol were 4.69 Å and 4.69 Å, respectively, and the equilibrium lattice constants of Au calculated using PBEsol and HSEsol were 4.08 Å and 4.09 Å, respectively. Hence, we used the lattice parameters calculated using PBEsol for the HSEsol calculations of electronic properties.

In addition to bulk Au$_2$O and Au$_2$O$_3$, *a*-surfaces (surfaces perpendicular to the *a*-axis) of Au$_2$O$_3$ in a vacuum were calculated using slab supercells. The vacuum width was 30 Å, which was sufficiently wider than the frequently used width (15 Å). For these supercells, the Monkhorst-Pack *k*-mesh [39] in the geometry relaxation was 1×3×3. A graphics-processing-unit acceleration [42, 43] was employed for calculations of Au$_2$O$_3$. Ion positions are displayed using VESTA [44].

**Table 1.** Lattice constants obtained from this study, experimental report [12], and previous DFT studies. The deviations of the theoretical values from the experimental values are shown in $\delta$.

| $Au_2O$ | This work | DFT [a] | DFT [b] | DFT [c] | DFT [d] |
|---|---|---|---|---|---|
| $a$ (Å) | 4.69 | 4.78 | 4.82 | 4.81 | 4.80 |

| $Au_2O_3$ | Experiment (Å) | This work (Å) | $\delta$ (%) | DFT [d] (Å) | $\delta$ (%) |
|---|---|---|---|---|---|
| $a$ (Å) | 3.838 | 3.906 | 1.77 | 4.038 | 5.21 |
| $b$ (Å) | 10.520 | 10.490 | 0.29 | 10.685 | 1.57 |
| $c$ (Å) | 12.827 | 12.840 | 0.10 | 13.057 | 1.79 |

[a] Ren et al. [23].    [b] Pei et al. [24].    [c] Filippetti and Fiorentini [18].    [d] Shi et al. [25].

### III. Results and Discussion
#### 1. Structural, elastic, and phonon properties

The only experimental data available for the physical properties of $Au_2O_3$ and $Au_2O$ pertain to crystal structures, bond lengths, and compositions determined by X-ray photoemission [10–15]. Therefore, the only experimental comparisons made in this study involve lattice constants, which are directly related to bond lengths. The computed equilibrium lattice constants are presented in Table 1. This table indicates that the lattice constants of $Au_2O$ and $Au_2O_3$ calculated using DFT (PBEsol) in this study are smaller than those reported in previous ones [18, 23–25].

Most GGAs are known to overestimate the lattice constants of solids [38], while calculations using PBEsol yield more accurate values for both lattice constants and ferroelectric properties [27, 29, 30, 38, 41]. The calculated lattice constants of $Au_2O_3$ in this study using PBEsol exhibited better agreement with experimental values than those reported previously [25]. The discrepancies between the present and earlier results arise from variations in the GGA functional used. Meanwhile, calculations performed using the conventional GGA (PBE) in this study yielded a lattice constant of 4.78 Å for $Au_2O$, consistent with prior reports [18, 23–25].

Table 2 presents the elastic compliance coefficients ($C_{ij}$) calculated using the GGA (PBEsol). These $C_{ij}$ coefficients are compared with those of well-known oxides, $SrTiO_3$ and $BaTiO_3$, since no literature on the elastic properties of Au oxides was found except for the bulk modulus of $Au_2O$ [24] and $C_{ij}$ of $Au_2O_3$ listed in Materials Project [45]. The $C_{ij}$ coefficients of $Au_2O$ for compression/expansion ($C_{11}$, $C_{12}$) are markedly smaller than those for $SrTiO_3$ and $BaTiO_3$. The bulk modulus of $Au_2O$, calculated [46] as 119 GPa using $C_{11}$ and $C_{12}$ from Table 2, is comparable to the previously reported value of 97 GPa [24]. Notably, the $C_{ij}$ coefficient for shear distortion ($C_{44}$) is almost zero, indicating an exceptionally soft response to shear stress. The $C_{ij}$ coefficients of $Au_2O_3$ for compression and expansion ($C_{ij}$: $i, j \leq 3$) are slightly smaller than those of $SrTiO_3$ and $BaTiO_3$, while the shear distortion coefficients ($C_{ii}$: $i \geq 4$) are substantially lower. These $C_{ii}$ values are slightly larger than those in Materials Project [45].

**Table 2.** Elastic compliance coefficients (GPa) obtained in the present study compared with those of cubic $SrTiO_3$ and tetragonal $BaTiO_3$ [31]. For comparison with Materials Project, the exchange of the indices 1–6 is needed.

|  | $C_{11}$ | $C_{12}$ | $C_{13}$ | $C_{22}$ | $C_{23}$ | $C_{33}$ | $C_{44}$ | $C_{55}$ | $C_{66}$ |
|---|---|---|---|---|---|---|---|---|---|
| $Au_2O$ | 123 | 117 |  |  |  |  | 3 |  |  |
| $Au_2O_3$ | 85 | 83 | 74 | 203 | 79 | 225 | 35 | 61 | 45 |
| $SrTiO_3$ | 349 | 103 |  |  |  |  | 114 |  |  |
| $BaTiO_3$ | 299 | 111 | 85 |  |  | 136 | 125 |  | 107 |

The calculated zone-center phonon modes of $Au_2O$ occur at energies of 76, 61, 26, 14, 10, and 6 meV (1 meV = 8.066 cm$^{-1}$ = 0.2418 Tz) and include three soft modes, whereas the previously reported energies were 70, 56, 25, 13, 10, and 6 meV [25]. Meanwhile, for $Au_2O_3$, the highest-energy zone-center phonon mode, at 72 meV [25], and phonon dispersions [45] are previously reported. We find that the zone-center phonon modes are grouped into "nearly continuous" bands, namely 61–75 meV, 39–44 meV, 36 meV, and 11–31 meV, and also include three soft modes. The first band comprises 48 modes: one at 75 meV, four at 74 meV, four at 73 meV, two at 72 meV, twelve at 71 meV, one at 70 meV, one at 68 meV, three at 67 meV, four at 66 meV, six at 65 meV, two at 64 meV, four at 63 meV, three at 62 meV, and one at 61 meV. The second band contains 15 modes: three at 44 meV, five at 43 meV, two at 42 meV, one at 40 meV, and four at 39 meV. The third band contains one mode at 36 meV. The fourth band contains 53 modes: two at 31 meV, three at 29 meV, six at 27 meV, six at 23–25 meV, sixteen at 20–22 meV, six at 17–19 meV, four at 16 meV, six at 12–13 meV, and four at 11 meV. These values are consistent with phonon dispersions in $Au_2O_3$ [45]. For example, in $SrTiO_3$ and $BaTiO_3$, the phonon energies calculated using PBEsol tend to agree better with the experimental values than those calculated using PBE [27].

Table 3. $E_g$ values obtained in this work (HSEsol) and previous studies a–e. a, b, d, e: GGA. c: screened-exchange LDA. In ref. [25], $E_F - E_V$, which should be zero, is approximately 0.4 eV for $Au_2O_3$.

|  | This work | DFT | DFT |
|---|---|---|---|
| $Au_2O$ | 0.96 | 0 [a] | 0 [b], 0.83 [c] |
| $Au_2O_3$ | 2.86 |  | 0.85 [d], 0.85 [e] |

[a] Pei et al. [24]. [b-d] Shi et al. [25]. [e] Materials Project [45].

## 2. Electronic structures

The band structures and density of states (DOS) were calculated using the hybrid functional HSEsol and the GGA-based DFT (PBEsol). Xiao et al. [26] reported a strong agreement between the $E_g$ values computed using hybrid functionals and through experiments for various semiconductors. To illustrate the accuracy of hybrid functionals for oxides, we compare the indirect band gaps obtained from experiments (exp.), hybrid functionals (HSE [22] or HSEsol [21]), and GGA-based DFT. Notably, the $E_g$ values calculated using HSE/HSEsol agree with experimental values, within approximately 0.2 eV. For cubic $SrTiO_3$, the $E_g$ values are 3.25 eV (exp.), 3.07 eV (HSE), and 1.80 eV (GGA); meanwhile, for tetragonal $BaTiO_3$, the $E_g$ values are 3.4 eV (exp.), 3.31 eV (HSEsol), and 1.76 eV (GGA) [27, 29].

The band diagrams and DOS of pure stoichiometric $Au_2O$ and $Au_2O_3$ are presented in Figs. 2 and 3. Throughout, all energies are measured relative to the Fermi level ($E_F$). Meanwhile, the $E_g$ values obtained in this and previous studies are compared in Table 3. In the *ab initio* results for 0 K, $E_F$ corresponds to the highest occupied states; for insulators, $E_F$ lies at the top of the valence band ($E_V$) [17,18,24,45], which differs from the definition of $E_F$ in semiconductor physics [1].

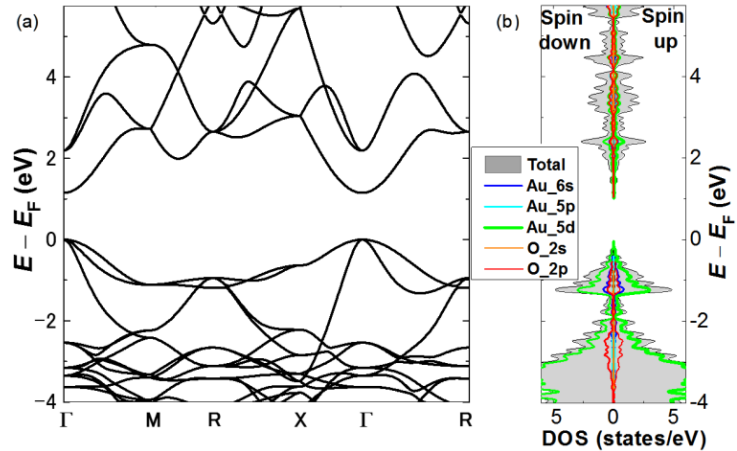

**Fig. 2.** Band structure (a) and projected DOS (b) of $Au_2O$ calculated using the hybrid functional (HSEsol).

The relative positions of the bands of $Au_2O$ were similar to those calculated using GGA [24] and the screened-exchange LDA [25]. For example, the band gap of $Au_2O$ occurs at the $\Gamma$ point and is a direct gap (Fig. 2(a)), agreeing with the previous reports [24, 25]. However, the calculated $E_g$ value is 0.96 eV, which is >20% greater than the values in these studies [24, 25] (Table 1). When calculated using the GGA (PBEsol), $E_g$ was zero, consistent with the calculations using GGA [24, 25]. In both $Au_2O$ and $Au_2O_3$, the DOS of both $Au_2O$ and $Au_2O_3$ consists primarily of $Au_{5d}$ and $O_{2p}$ orbitals in the energy range from −4 eV to 5 eV, and magnetic ordering is absent, as the projected DOSs and band diagrams are identical for their spin-up and spin-down states (Figs. 2(b) and 3(b)).

In Fig. 3(b), the DOSs of $Au_2O_3$ are divided by four because the number of Au atoms in $Au_2O_3$ is four times that in $Au_2O$. The band gap is indirect, consistent with previous studies [25,45]. However, $E_g$ is 2.86 eV (Figs. 3(a) and 3(b)), which suggests that a pure, stoichiometric $Au_2O_3$ crystal is transparent and nearly colorless. Contrastingly, Shi et al. [25] described $Au_2O_3$ as a semiconductor with $E_g$ = 0.85 eV (Table 3), while the DOS in the same study [25] indicates $E_g \approx$ 0.5 eV or less. When calculated using the GGA (PBEsol), the $E_g$ value was 0.87 eV, consistent with Materials Project [45].

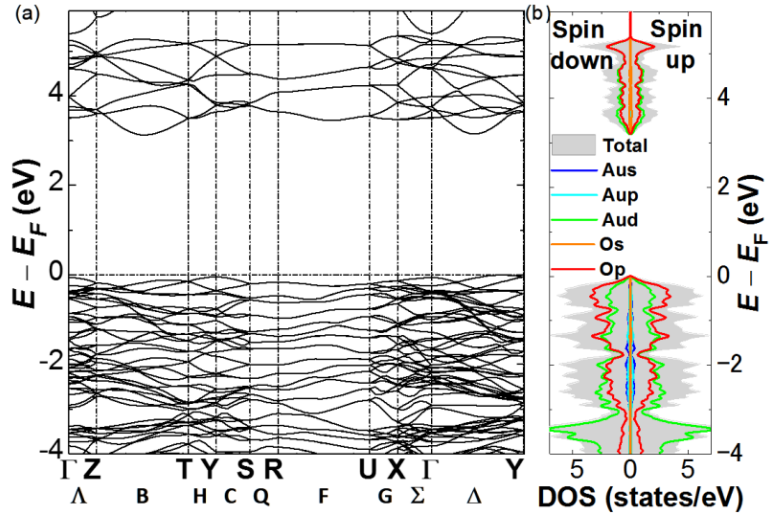

**Fig. 3.** Band structure (a) and projected DOS (b) of $Au_2O_3$ calculated using the hybrid functional (HSEsol). The original DOS is divided by four to obtain the displayed DOS to adjust the number of Au atoms in the unit cell (4 Au atoms in $Au_2O$ and 16 Au atoms in $Au_2O_3$).

In the detailed analysis of Fig. 3(a), many bands are closely packed due to low structural symmetry, as previously reported [25,45]. Consequently, several points in the Brillouin zone [47] exist near $E_V$ (i.e., $E_F$) and the bottom of the conduction bands ($E_C$). Specifically, $E_V$ is on the Γ–Y path, with the energies at Γ and Y close to $E_V$, while $E_C$ lies on the T–Z path, with the energies on the Γ–Y path close to $E_C$. Although not shown in Fig. 3(a), the energies on the Γ–R, Γ–Z, and X–Y paths were also close to $E_V$, and the energies on the Γ–R and Γ–Z were also close to $E_C$, which was also found in the calculations using PBEsol.

Figure 3(a) can be favorably compared with the bands in Materials Project [45], using the high symmetry points present in both plots. For this comparison, we denote $E_Γ$ and $E_U$ as the highest energy at Γ below $E_V$ and the lowest energy at U above $E_C$, respectively. $E_F − E_Γ$ is 70 meV in Fig. 3(a) and 130 meV in ref. [45], which can be considered as consistent with DOS. In Fig. 3(a), $E_C$ lies on the T–Z path and is 0.42 eV lower than $E_U$, whereas in ref. [45], the T–Z path is absent, and $E_C$ is only 0.14 eV lower than $E_U$. Hence, $E_C$ as compared with $E_U$ in Fig. 3(a) is much lower than $E_C$ as compared with $E_U$ in ref. [45], which presumably suggests that the band paths in conduction bands in Fig. 3(a) are approximately sufficient for estimating $E_C$.

A comparison between these results and those in ref. [25] is unfeasible because the band plots in Fig. 3(a) of ref. [25] are incomplete; specifically, the bands beneath $E_F$, which DOS in that figure indicates should be present, are missing. In that figure, the standard convention for $E_F$ in insulators ($E_F = E_V$) appears to be applied, as $E_F = E_V$ in Fig. 4 of the same study [25]. Hence, the bands corresponding to the correct $E_V$ are absent in Fig. 3(a) of ref. [25], which explains the discrepancy between the reported $E_g = 0.85$ eV [25], presumably obtained using $E_C − E_V$ and an incorrect $E_V$, and $E_g ≈ 0.5$ eV calculated using $E_C − E_F$ or DOS in Fig. 3(a) of ref. [25].

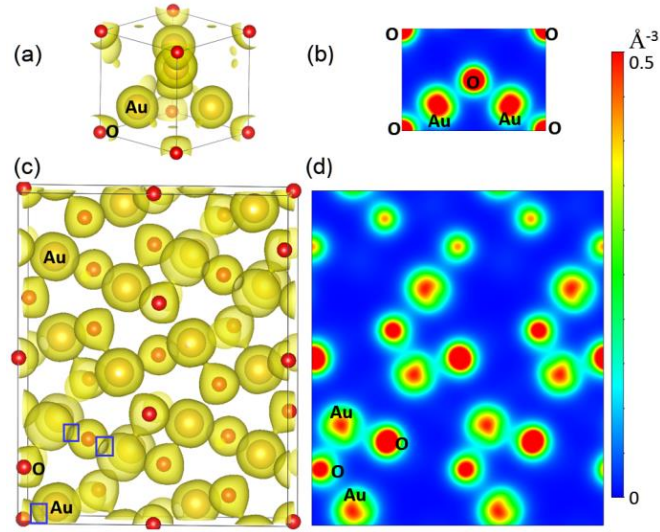

**Fig. 4** Iso-$e^-$-density surfaces (0.12 Å$^{-3}$) in Au$_2$O (a) and Au$_2$O$_3$ (c) and their cross-sections at the (110) plane ((b), (d)). The ovals at the boundaries of the O and Au orbits in (c) indicate that these orbits are in contact there, with examples marked by blue squares.

The greater stability of Au$_2$O$_3$ compared to Au$_2$O [6, 16, 17] indicates stronger bonding, i.e., increased hybridization between Au and O in Au$_2$O$_3$. This is confirmed by the higher DOS and the stronger correlation between the Au$_{5d}$ and O$_{2p}$ orbitals just beneath $E_F$ in Au$_2$O$_3$ compared to Au$_2$O, as shown in the DOS plots in Figs. 2(b) and 3(b). Consistently, many Au and O iso-$e^-$-density surfaces are in contact in Au$_2$O$_3$ in Fig. 4, whereas those in Au$_2$O are not, supporting the stronger bonding between Au and O atoms in Au$_2$O$_3$. Furthermore, the atomic bonding (Fig. 1) shows that each Au atom is coordinated three-dimensionally with four O atoms in Au$_2$O$_3$ but one-dimensionally with only two O atoms in Au$_2$O. In particular, Au$_2$O consists solely of a $V$-shaped and an inverted-$V$-shaped Au-O bonds, their vertices contacted at an O atom (Figs. 1(a) and 4(a)). This bonding configuration is speculated to be rotationally deformable. Contrastingly, the bonding in SrTiO$_3$ and BaTiO$_3$ is three-dimensional, and the Ti$_{3d}$ orbits are highly directional. These features may explain the aforementioned supersoftness of Au$_2$O under shear strain.

The synthesis of Au$_2$O suitable for electrical or optical characterization has not yet been reported. Similarly, the synthesis of extremely stoichiometric Au$_2$O$_3$ remains challenging [13]; e.g., vacuum-deposited Au$_2$O$_3$ films were Au$_2$O$_{3-x}$ ($x$ = 0.17) [15], whereas the theoretical studies of Au$_2$O$_{3-x}$ ($x$ > 0) have not been reported. In light of these limitations, we present the DOS of Au$_2$O$_{3-x}$ ($x$ = 4%, 8%), with the positions of the oxygen vacancies (V$_o$) indicated in Fig. 5.

The presence of $V_o$ did not induce metallicity in $Au_2O_3$ (Fig. 5). However, a single $V_o$ among 24 oxygen atoms ($x = 4\%$) lowers $E_C$ to 1.5 eV, while two $V_o$ species ($x = 8\%$) lower $E_C$ to 1.3 eV. These results suggest that $Au_2O_3$ containing $V_o$ species is expected to be a semiconductor and appear nearly black. The configuration with $V_o$ species at positions 1 and 2 corresponds to a dispersive $V_o$ distribution, whereas that with $V_o$ species at positions 1 and 3 corresponds to an aggregated distribution. Notably, the free energy gain, i.e., the stability of the aggregated configuration is 0.2 eV per unit cell higher than that of the dispersive configuration.

$V_o$'s in $Au_2O_3$ create trap-like narrow bands near the original $E_V$ and $E_C$, while both the valence and conduction bands near $E_F$ and these narrow bands exhibit a mixture of $O_{2p}$ and $Au_{5d}$ character with nearly equal contributions. The $V_o$-related new band near $E_V$ is ineffective as acceptors because it lies below $E_F$, and the $V_o$-related new bands near $E_C$ are also ineffective as donors because they are located well above $E_F$. This behavior contrasts notably with that observed in Ti-oxides such as $SrTiO_3$, where the valence and conduction bands near $E_F$ are predominantly of $O_{2p}$ and $Ti_{3d}$ character, respectively. In these oxides, the $V_o$-created donor states are of $Ti_{3d}$ character and lie slightly below $E_C$, and the trapped $e^-$ in these shallow states can be easily activated to become free $e^-$, which explains experimental properties well.

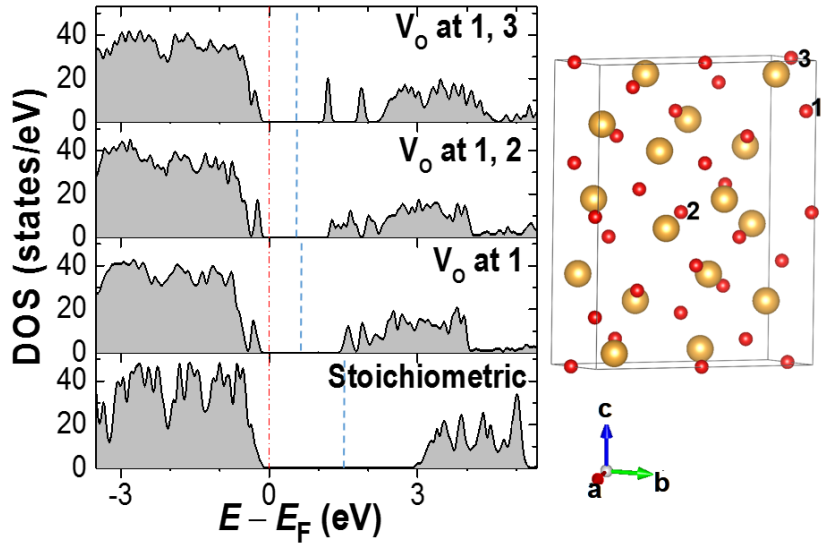

**Fig. 5.** DOS of $Au_2O_3$ with 4% and 8% oxygen vacancies (Vo) calculated with hybrid functional and the illustration of the position of Vo. The vertical red and blue lines show $E_F$ in *ab initio* calculations for insulators and semiconductor physics, respectively [1].

### 3. $P_S$ and dielectric properties

The electronic dielectric constant ($\varepsilon$), or the optical dielectric constant at zero energy, of $Au_2O$ is 10.6 (Table 4), which is slightly higher than the previously reported value of $\varepsilon = 8.5$ [24]. The $\varepsilon$ values of $Au_2O_3$ are comparable to those reported in Materials Project [45]. These $\varepsilon$ values are considerably higher than those of the archetypal ferroelectric tetragonal $BaTiO_3$ and the archetypal paraelectric cubic $SrTiO_3$ ([48]), both of which are recognized as high-permittivity materials (Table 4).

Furthermore, the total dielectric constants under ion-clamped conditions are comparable to those of $BaTiO_3$ and $SrTiO_3$: 11 for $Au_2O$; 7, 12, and 11 for $Au_2O_3$; 7 and 28 for $BaTiO_3$; and 12 for $SrTiO_3$.

**Table 4.** $P_S$ (μC/cm$^2$) and $\varepsilon$ along the $a$-, $b$-, and $c$-axes obtained in the present study, compared with values from previous studies. All data were calculated using GGA(PBEsol) except for 5.7 $^{c*}$, which was calculated with DFT with Hubbard on-site potential [30] and agreed well with the experimental data at optical energy.

|   | $P_{Sa}$ | $P_{Sb}$ | $P_{Sc}$ | $\varepsilon_a$ | $\varepsilon_b$ | $\varepsilon_c$ | $\varepsilon_a$ | $\varepsilon_c$ |
|---|---|---|---|---|---|---|---|---|
| $Au_2O$ | 0 | 0 | 0 | 10.6 | | | 8.5 [a] | |
| $Au_2O_3$ | 213.8 | 0 | 0 | 6.6 | 11.0 | 9.9 | | |
| $BaTiO_3$ | 0 | 0 | 32.8 [b] | | | | 6.5 [c] | 5.8 [c] |
| $SrTiO_3$ | | | | | | | 6.3 [c], 5.7 [c*] | |

[a] Ref. [24]. [b] Ref. [30]. [c] Ref. [48].

We observed that the piezoelectric constants are zero for $Au_2O$ but nonzero for $Au_2O_3$, consistent with the absence of inversion symmetry in the latter. Likewise, $P_S$ is zero for $Au_2O$ but nonzero for $Au_2O_3$ (Table 4). As demonstrated in the comparison with $BaTiO_3$ in Table 4, the $P_S$ of $Au_2O_3$ is substantial.

For reference, the experimental $P_S$ of $BaTiO_3$ at 270 K is 26.9 μC/cm$^2$, while the quasi-experimental $P_S$ of $BaTiO_3$ at 0 K is 32μC/cm$^2$ [29], which serves as the appropriate benchmark for comparison with *ab initio* calculations [29]. Further, the $P_S$ of $BaTiO_3$ reported in Table 4 agrees well with this reference $P_S$. In this context, recent outcomes [31, 32] report that potential produced by $P_S$ can enhance photocatalysis when defect-free surfaces form without adsorbates. Furthermore, we speculate that $Au_2O_3$-like atomic configurations may locally exist on oxygen-adsorbed Au surfaces and thereby contribute to catalytic activity.

These applications utilize free polar surfaces, to which $P_S$ is perpendicular. In this case, $P_S$ is shown to disappear or be substantially reduced by the depolarization field ($E_d$) arising from $P_S$ [49–51]. This situation corresponds to a prototypical case of a ferroelectric wherein a freestanding ferroelectric is plate-like, placed in an infinite vacuum, free of adsorbates and defects, and of the single-domain with $P_S$ perpendicular to the plate surface. In this case, $E_d = -P_S/\varepsilon_0$ ($\varepsilon_0$: vacuum permittivity) and produces a macroscopic equilibrium potential difference $\Delta\phi^{eq}_{macro}$, when the ferroelectric is perfectly insulating. No additional permittivity should be inserted in $E_d = -P_S/\varepsilon_0$, when $P_S$ stands for an all-inclusive or self-consistent spontaneous polarization [49].

In the Ginzburg–Landau (GL) framework for this case, the effect of $E_d$ is expressed by an effective increase in the ambient temperature ($T$) [49]. The GL free energy is

$$F = (T-T_0)P^2/2C\varepsilon_0 + \beta P^4/4 + \gamma P^6/4 - PE_d/2,$$

where $T_0$, $C$, $\beta$, and $\gamma$ are Curie-Weiss temperature, Curie constant, and GL coefficients, respectively. Curie temperature $T_C$ is $T_0 + \Delta T$, where $\Delta T = 3\beta^2/16\gamma C\varepsilon_0$. For 2nd order transition, $\gamma = 0$, $\beta > 0$, and $T_C = T_0$. The effect of $E_d$ is the change of $T_0$ to $T_0 - C$ in $F$, or to $T_0 - C/\varepsilon_{NG}$ when the inaccuracy of the GL theory is corrected by a factor of $\varepsilon_{NG} > 1$ [49]. For almost all displacive-type ferroelectrics such as $BaTiO_3$ and $PbTiO_3$, $T_0 \ll C, C/\varepsilon_{NG}$, which means $P_S = 0$ and the emergence of an inversion symmetric phase in this situation.

Consequently, the previous studies [49–51] have demonstrated that, in the aforementioned situation, an insulating ferroelectric without external screening charges either transforms into a paraelectric phase (i.e., $P_S = 0$) or remains ferroelectric (i.e., $P_S \neq 0$), with the latter accompanied by the formation of $e^-$ and $h^+$ layers. Hence, studies [50,51] have shown that the maximum $\Delta\phi^{eq}_{macro}$ ($\Delta\phi^{eq}_{macro}{}^{max}$) generated by $P_S$, corresponding to $\Delta\phi^{eq}_{macro}$ between the $P_S^+$ and $P_S^-$ faces, is less than $E_g/e$, where $e$ is the elementary charge.

For instance, both DFT and GL theories show that a freestanding $BaTiO_3$ sample becomes paraelectric below a thickness of 100 Å but remains ferroelectric above this threshold, forming an $e^-$ layer on the $P_S^+$ surface and a $h^+$ layer on the $P_S^-$ surface [49–51]. Similar $e^-$ and $h^+$ layers have also been suggested to exist at charged domain boundaries [51]. A comparable phenomenon has also been theoretically observed in the hyper-ferroelectric LiBeSb, which remains ferroelectric even at a thickness of 6.08 Å when freestanding in vacuum and features surface $e^-$ and $h^+$ layers [48].

### 4. *a*-face $Au_2O_3$ nanosheets

The above theoretical studies on ferroelectrics [49–51] suggest that placing a freestanding nano-ferroelectric in vacuum can induce a paraelectric phase due to $E_d$, implying the existence of $T_C$ similar to ferroelectrics. Here, the polar surfaces of $Au_2O_3$ are *a*-faces.

In light of this, we explore the same possibility in $Au_2O_3$, although its unit cell differs substantially from those of conventional displacive-type ferroelectrics. To this end, we examined freestanding *a*-face $Au_2O_3$ nanosheets with thicknesses of one, two, and five unit cells with a 30-Å-wide vacuum. These values correspond to physical thicknesses of 3.91 Å, 7.81 Å, and 19.4 Å, respectively. Here, $\Delta\phi^{eq}_{macro}{}^{max}$ represents the built-in potential difference across the vacuum between the top and bottom surfaces.

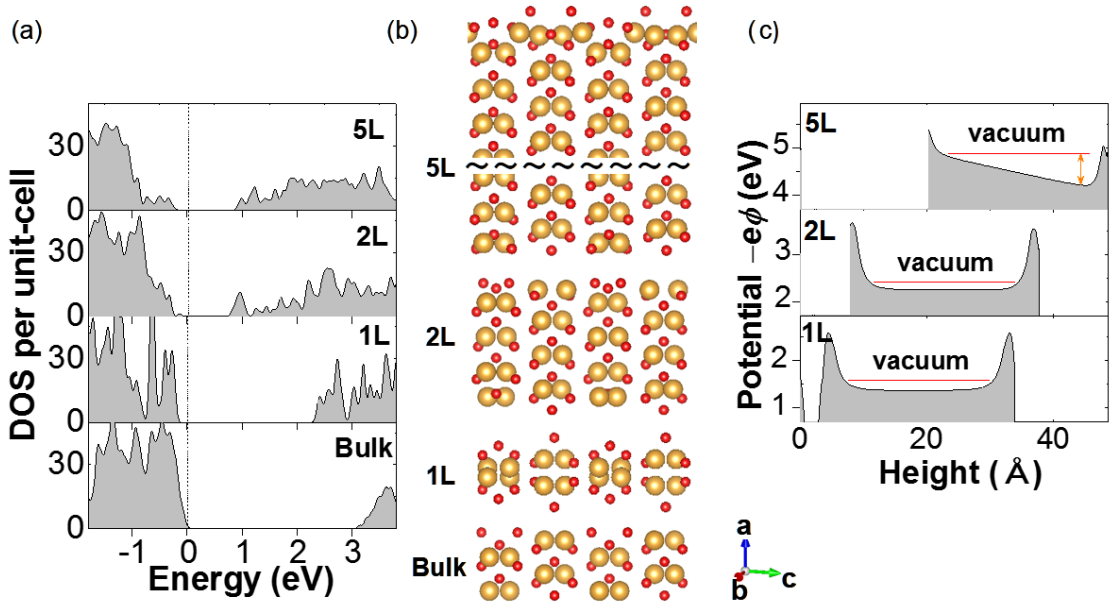

**Fig. 6.** DOSs of bulk and *a*-face $Au_2O_3$ nanosheets with thicknesses of 1, 2, and 5 unit cells (L) in vacuum (a), atomic positions (b), and the corresponding potential in vacuum (c). The unit layer of bulk consists of two sub-unit layers that differ only slightly from each other. When $P_S$ is zero and the structure is symmetric along *a*-axis, the potential in vacuum is constant; the horizontal orange lines show a constant potential.

All the examined nanosheets are found to be insulating (Fig. 6(a)). Hence, $\Delta\phi^{eq}_{macro}{}^{max} = -\int dz\, E_d(z) = \int dz\, P(z)/\varepsilon_0$, where the integration is from the bottom to top surfaces along $a$-axis, and $P$ is the total polarization including $P_S$. For macroscopic systems that ignore surface atomic contributions, $P(z) = P_S$. Because of the periodic boundary condition, this $\Delta\phi^{eq}_{macro}{}^{max}$ yields $-\Delta\phi^{eq}_{macro}{}^{max}$ in the vacuum. Hence, $\Delta\phi^{eq}_{macro}{}^{max} = 0$ means that an average $P$ or a net $P_S$ is zero.

In Fig. 6(b), buckling of the Au–O bonds—that is, the outward displacement of O atoms relative to Au atoms—at the surfaces is evident, consistent with that observed in other oxides [52]. However, the image of the five-unit-cell-thick nanosheet reveals that both Au and O atoms are substantially displaced at the top ($P+$) surface, causing the original unit cell to be completely disrupted. In contrast, the relative atomic positions remain nearly unchanged within the inner region, which is located between one unit-cell below the top and half a unit cell above the bottom and has a thickness of 13.4 Å.

Table 5 Inversion symmetry along the $a$-axis in atom positions of the one-unit-cell-thick $Au_2O_3$ (having 16 Au and 24 O atoms): comparison of the $a$-axis coordinates of two Au ($a_{Au}$) and two O ions ($a_O$) at each $c$-axis coordinate ($c_{Au}$) in Å. The centers of mass of $a_{Au}$ ($a_{Au}{}^{av}$) and that of $a_O$ ($a_O{}^{av}$) matched to five decimal places. $a_{Au}$ and $a_O$ listed below are obtained by subtracting $a_O{}^{av}$ from the original data.

| $c_{Au}$ | 0.640 | 2.606 | 3.814 | 5.780 | 7.059 | 9.025 | 10.23 | 12.199 |
|---|---|---|---|---|---|---|---|---|
| $a_{Au}$ | 0.2828 | −0.7596 | 0.7602 | −0.2830 | 0.2828 | −0.7596 | 0.7602 | −0.2830 |
| $a_{Au}$ | −0.2830 | 0.7602 | −0.7596 | 0.2828 | −0.2830 | 0.7602 | −0.7596 | 0.2828 |

| $c_O$ | 0 | 0.925 | 2.075 | 3.210 | 4.344 | 5.495 |
|---|---|---|---|---|---|---|
| $a_O$ | −1.5283 | −0.9909 | 0.5508 | 1.9972 | −0.5503 | 0.9907 |
| $a_O$ | 1.5278 | 0.9907 | −0.5503 | −1.9968 | 0.5508 | −0.9909 |

| $c_O$ | 6.420 | 7.344 | 8.495 | 9.629 | 10.764 | 11.914 |
|---|---|---|---|---|---|---|
| $a_O$ | 1.5278 | −0.9909 | 0.5508 | −1.9968 | −0.5503 | 0.9907 |
| $a_O$ | −1.5283 | 0.9907 | −0.5503 | 1.9972 | 0.5508 | −0.9909 |

In the one-unit-cell-thick nanosheet, Au and O ion positions substantially changed, resulting in inversion symmetry along the $a$-axis, i.e., the original $P_S$ direction as demonstrated in Fig. 6(b) and Table 5, consistent with the theories mentioned above [48–50]. Accordingly, the nanosheet loses net $P_S$, as indicated by $\Delta\phi^{eq}_{macro}{}^{max} = 0$ (Fig. 6(c)) as predicted [48–50]; an inversion symmetric phase can be found by placing a three-dimensional nanoscale ferroelectric in vacuum.

The $\Delta\phi^{eq}_{macro}$ profiles in the one- and five-unit-cell-thick nanosheets in Fig. 5(c) were very similar to those in the nanosheets of prototypical ferroelectrics, which show $P_S = 0$ in thin sheets and $P_S \neq 0$ in thick sheets [49], while Fig. 5(c) displays $\Delta\phi^{eq}_{macro}$ only in vacuum to facilitate the comparisons.

$\Delta\phi^{eq}_{macro}{}^{max} \neq 0$ in the five-unit-cell-thick nanosheet in Fig. 6(c) indicates the presence of $P_S$. Furthermore, as mentioned above, the atomic positions in the inner region of the five-unit-cell-thick nanosheet remain nearly identical to those in bulk $Au_2O_3$ (Fig. 6(b)). This suggests that $P_S$ in the inner region of the nanosheet is similar to that of bulk $Au_2O_3$. Therefore, we approximate that the bulk $P_S$ value is preserved in the inner region. In this case, $\Delta\phi^{eq}_{macro}{}^{max}$ is 320 V (= 13.4 Å $\times P_S/\varepsilon_0$) in the absence of any potential screening mechanisms [48]. Similarly, the top surface ($P+$ surface) of a two-unit-cell-thick nanosheet also exhibits evident displacements of O atoms (Fig. 6(b)).

In contrast, the computed $\Delta\phi^{eq}_{macro}{}^{max}$ is 0.6 V in Fig. 6(c) and less than $E_g/e$, consistent with theoretical predictions for conventional ferroelectrics [49–51]. Notably, this reduction in $\Delta\phi^{eq}_{macro}{}^{max}$ for the two- and five-unit-cell-thick $Au_2O_3$ nanosheets primarily results from surface atomic displacements (reconstructions), unlike in displacive-type ferroelectrics such as $PbTiO_3$ and $BaTiO_3$. Given the thickness dependence of $\Delta\phi^{eq}_{macro}{}^{max}$ shown in Fig. 6(c) and the occurrence of structural changes only at the surfaces, we conjecture that the built-in potential also exists in thicker sheets and is likely greater than 0.6 eV. The presence of $P_S$ in a freestanding 19.4 Å-thick-nanosheet and $T_C$ suggests that $Au_2O_3$ can be classified as hyper-ferroelectric-like.

Conventionally, the screening mechanism of $E_d$ in ferroelectrics without intentional screening materials, such as electrodes, is primarily due to defects and the $e^-$ and $h^+$ supplied from defects. In ideally prepared ferroelectrics, intrinsic band-excited $e^-$ and $h^+$ can provide screening [50]. Although often misunderstood, buckling at surfaces [52] is insufficient to screen $E_d$ because it enhances $P_S$ at the $P-$ surface. From this perspective, the substantial screening found in $Au_2O_3$ reveals a new class of screening mechanism in defect-free insulators.

**IV. Conclusion**

Because Au oxides form under various application conditions [2–4, 53–59], understanding their properties is essential for advancing related technologies, while bulk forms have been experimentally observed only in $Au_2O_3$ and $Au_2O$ [10–15]. Among these, merely $Au_2O_3$ is stable under ambient conditions [16,17], can exist as single crystals [11], and appears in various experiments and applications [53–59]. However, the properties of these oxides, especially $Au_2O_3$, have not been much clarified theoretically [25,45]. Hence, this study examined the structural, electronic, dielectric, elastic, and phonon properties of $Au_2O_3$ and $Au_2O$, using DFT (PBEsol) and a hybrid functional to ensure accuracy. As benchmark tests, the calculated lattice parameters showed improved agreement with available experimental values compared to previous reports.

These calculations revealed a high degree of functionality in these oxides, including pronounced $P_S$ along the *a*-axis in $Au_2O_3$, which may enhance catalytic activity [31,32]. $P_S$ was retained even in 2-nm-thick nanosheets, yielding a potential of 0.6 V. Because this reduction in potential was caused by the heavy surface reconstruction and the potential increased with thickness (Fig. 6(c)), we conjecture that similar $P_S$ and potentials of comparable magnitudes are present in thicker sheets. Contrastingly, one- and two-unit-cell-thick $Au_2O_3$ nanosheets lost $P_S$, and the one-unit-cell-thick nanosheet exhibited inversion symmetry along the *a*-axis, i.e., the original $P_S$ direction, suggesting a new approach to finding a paraelectric phase. The substantial screening of $E_d$ by heavy surface reconstruction, which has not been reported in typical ferroelectrics, is considered a novel class of screening mechanisms in ideally prepared insulators.

Previous theories have identified $Au_2O_3$ as a semiconductor with an $E_g$ of 0.85 eV [25,45], which remains a standard to date. The present hybrid functional calculations revealed that $Au_2O_3$ was an insulator with $E_g$ = 2.86 eV. These calculations also indicated that the minimum band gap of stoichiometric $Au_2O$ occurred at the $\Gamma$ point and was 0.96 eV, 20% greater than the previous theoretical value [25]. The $E_g$ values estimated using PBEsol (GGA) were in agreement with those previously calculated using GGA [24, 25]. The effect of $V_O$ in $Au_2O_3$ was notably different from that in Ti-oxides, and $V_O$ changed $Au_2O_3$ into a semiconductor with $E_g$ of 1.3–1.5 eV but retained insulativity.

Finally, $Au_2O$ exhibited unusual softness under shear distortion. Both $Au_2O$ and $Au_2O_3$

displayed high dielectric constants. DFT and hybrid functional theory calculations revealed no magnetization in stoichiometric Au$_2$O and Au$_2$O$_3$.

**Declaration of Conflicts of Interest**     The authors declare no competing interests.

**Declaration of generative AI use**     No generative AI was used in this study.

**Data availability** The data in our manuscript will be obtained from the corresponding author upon request.

**Acknowledgments** This work was supported partly by the Murata Science and Education Foundation. The discussions with Peter Blöchel of TU Clausthal and Kaushik Dayal of Carnegie Mellon Univ. are greatly acknowledged.